\documentclass{elsart}%
\usepackage{amsfonts}
\usepackage{amsmath}
\usepackage{amssymb}
\usepackage{graphicx}%
\providecommand{\U}[1]{\protect\rule{.1in}{.1in}}
\begin{document}
\begin{frontmatter}
\title{Ultra High Energy Cosmic Rays from decay of Holeums in Galactic Halos}
\author{Abhijit L. Chavda$^{a}$\thanksref{corr1}, L. K. Chavda$^{b}$}
\address[lngs]{Physics Department, V.N. South Gujarat University, Udhna-Magdalla Road, Surat - 395007, Gujarat, India.}
\thanks[corr1]{E-mails:\\a01.l.chavda@gmail.com\\holeum@gmail.com}
\address[AQ]{49, Gandhi Society, City Light Road, Via Parle Point, Surat - 395007, Gujarat, India}
\begin{abstract}
Stable, quantized gravitational bound states of primordial black holes called Holeums could have been produced in the early universe and could be a component of the Super Heavy Dark Matter (SHDM) present in galactic halos. We show that Holeums of masses $\sim$ $10^{13}$ $-$ $10^{14}$ GeV and above are stable enough to survive in the present-day universe. We identify such Holeums as promising candidates for the SHDM "X-particle" and show that the decay of such Holeums by pressure ionization can give rise to cosmic rays of all observed energies, including UHECR. The absence of the GZK cut-off is explained by the galactic halo origin of the UHECR. We predict that the cosmic rays are a manifestation of the end-stage Hawking radiation burst of the primordial black holes (PBH) liberated by the ionization of Holeums. Antimatter detected in cosmic rays could be a signature of their Holeum origin.
\end{abstract}
\end{frontmatter}

\section{Introduction}

The origin of the Ultra High Energy Cosmic Rays (UHECR) is one of the most
profound mysteries in astrophysics
\cite{F.Stecker,S.Sarkar,P.Bhattacharjee.et.al}. Cosmic ray particles with
energies exceeding $10^{20}$ eV have been observed
\cite{J.Linsley,R.Brownlee.et.al,M.Winn.et.al,A.Watson,D.Bird.et.al,N.Hayashida.et.al,S.Yoshida.et.al,M.Takeda.et.al}%
, which has been interpreted by some to indicate the existence of Zevatrons in
the universe \cite{R.Blandford,T.W.Jones,M.Honda.et.al,E.Armengaud.et.al},
which are sources that accelerate cosmic ray particles to energies as high as
one ZeV = $10^{21}$ eV. There is much debate about the origin of UHECR
\cite{M.Nagano.et.al,M.Nagano2,J.Krizmanic.et.al}.

The prevailing theories of the origin of UHECR fall into two categories:
"bottom-up" acceleration \cite{P.Bhattacharjee.et.al}, and "top-down" decay
\cite{A.G.Doroshkevich.et.al,M.Drees}. In the bottom-up scenario, particles
are accelerated from low energies to ultra-high energies in certain special
astrophysical environments. Examples are: acceleration in shocks associated
with supernova explosions \cite{Stanev.et.al}, active galactic nuclei (AGNs)
\cite{A.P.Szabo.et.al,V.Berezinsky.et.al} and radio lobes of powerful radio
galaxies \cite{Biermann.et.al}. In the top-down scenario, on the other hand,
UHECR originate from the decay of certain sufficiently massive particles
originating from physical processes in the early Universe, and no acceleration
mechanism is needed. Examples are: Topological Defects
\cite{N.A.Porter,V.Berezinsky.et.al2,V.Berezinsky2} and Super Heavy Dark
Matter (SHDM) \cite{A.G.Doroshkevich.et.al,C.Barbot}, among many others.

Most of the currently favoured candidates for sources of UHECR, such as radio
galaxies \cite{F.Takahara,P.Biermann,P.Biermann2,R.Protheroe}, etc. lie at
large cosmological distances ($\gg$ $100$ Mpc) from the earth. This is a
stumbling block if the UHECR particles are hadrons or heavy nuclei, due to the
Greisen-Zatsepin-Kuzmin (GZK) effect \cite{K.Greisen,G.Zatsepin.et.al}.
Greisen and, independently, Zatsepin and Kuzmin noted that cosmic ray nucleons
with energies exceeding $\sim$ $4\times10^{10}$ GeV would suffer severe energy
losses through scattering on the Cosmic Microwave Background (CMB) photons due
to photo-production of pions. The mean-free path for the CMB photons is of the
order of a few Mpc \cite{F.Stecker2}. Consequently the typical range of cosmic
ray nucleons should decrease rapidly above this energy, leading to a `GZK
cut-off' in the energy spectrum if the sources are cosmologically distant. The
absence of such a cut-off would imply that the sources are nearby, within the
local supercluster. Recent calculations \cite{T.Stanev.et.al}\ indicate that
the typical range of a nucleon drops below $\sim100$ Mpc above $10^{11}$ GeV.
The mean free path is estimated to be $\sim1-5$ Mpc at $10^{11}$ GeV
\cite{R.Protheroe.et.al,R.Protheroe.et.al.2,R.Protheroe.et.al.3,R.Protheroe.et.al.4}%
. It therefore appears unlikely that UHECR would originate from large
cosmological distances.

In this paper, we put forth the Holeum \cite{Chavda.Chavda} as a SHDM
"X-particle" and demonstrate that the ionization of Holeums in galactic halos
can give rise to cosmic rays of all observed energies, including UHECR. The
absence of the GZK cut-off can be attributed to the galactic halo origin of
the UHECR. The remarkable feature of this model is that the cosmic rays are a
manifestation of the end-stage Hawking radiation burst of the primordial black
holes (PBH) liberated in ionizing collisions among the stable Holeums in
galactic halos. The Hawking radiation origin can explain the detection of
antimatter in cosmic rays.

\section{The Holeum as a SHDM X-particle}

\subsection{Background}

Over the last 20 years, compelling theoretical and observational evidence has
built up that the visible matter in the universe makes up a very small
fraction of the total matter composition of the universe \cite{K.A.Olive}. It
is now accepted that the universe is composed of approximately 73\% dark
energy, 23\% dark matter, and about 4\% visible matter
\cite{M.J.Rees,D.B.Cline,W.Boer}. Consequently, the nature and identity of the
dark matter of the Universe has become one of the most challenging problems
facing modern cosmology. Several candidates for dark matter particles have
been proposed, which include Standard Model neutrinos
\cite{G.Bertone.et.al,L.Bergstrom}, Sterile neutrinos
\cite{S.Dodelson.et.al,G.Bertone.et.al}, Axions
\cite{G.Bertone.et.al,P.Sikivie2}, Supersymmetric particles
\cite{G.Bertone.et.al,T.Falk.et.al,J.L.Feng.et.al,K.Choi.et.al,L.Covi.et.al,L.Covi.et.al2}%
, dark matter from Little Higgs models
\cite{N.Arkani-Hamed.et.al,N.Arkani-Hamed.et.al2,N.Arkani-Hamed.et.al3,N.Arkani-Hamed.et.al4,H.Cheng.et.al}%
, Kaluza-Klein dark matter \cite{H.Cheng.et.al2,A.Barrau.et.al.2}, WIMPZILLAs
\cite{E.Kolb.et.al}, Cryptons \cite{J.R.Ellis.et.al,J.R.Ellis.et.al.2},
primordial black holes \cite{A.Dolgov.et.al,D.Blais.et.al,N.Ashfordi.et.al},
WIMPs \cite{R.Bernabei.et.al}, and super-heavy X-particles
\cite{C.Barbot,C.Barbot.et.al,C.Barbot.et.al.2,Sanghyeon.Chang.et.al,Karim.Benakli.et.al,K.Hamaguchi.et.al,K.Hamaguchi.et.al2,K.Hamaguchi.et.al3,C.Coriano.et.al,D.J.Chung.et.al,V.Berezinsky}%
. Galaxies are observed to have dark matter halos which contain far more
matter than their visible regions \cite{F.Combes,H.Hoekstra.et.al,A.Burkert}.
Most galaxies are not dominated by dark matter inside their optical disks
\cite{F.Combes,H.Hoekstra.et.al,A.Burkert}, which suggests that the dark
matter has properties that segregate it from visible matter.

It has been theorized that super-heavy X-particles
\cite{C.Barbot,C.Barbot.et.al,C.Barbot.et.al.2,Sanghyeon.Chang.et.al,Karim.Benakli.et.al,K.Hamaguchi.et.al,K.Hamaguchi.et.al2,K.Hamaguchi.et.al3,C.Coriano.et.al,D.J.Chung.et.al,V.Berezinsky}
left over as relics from the primordial universe may be a constituent of the
dark matter present in the universe. These X-particles are believed to have
masses in the range $M_{X}\sim10^{12}-10^{16}$ GeV and are expected to exhibit
the properties that characterize the matter contained in the dark matter halos
of galaxies.

\subsection{Properties of Holeum}

The theory of Holeum was presented in \cite{Chavda.Chavda}. The energy
eigenvalue $E_{n}$ of a Holeum consisting of two identical primordial black
holes of mass $m$ is given by%
\begin{equation}
E_{n}=-\frac{mc^{2}\alpha_{g}^{2}}{4n^{2}} \label{1}%
\end{equation}
where $n$ is the principal quantum number, $n=1,2,...\infty$ and $\alpha_{g}$
is the gravitational analogue of the fine structure constant, given by%
\begin{equation}
\alpha_{g}=\frac{m^{2}G}{\hbar c}=\frac{m^{2}}{m_{P}^{2}} \label{2}%
\end{equation}
where%
\begin{equation}
m_{P}=\left(  \frac{\hbar c}{G}\right)  ^{\frac{1}{2}} \label{3}%
\end{equation}
is the Planck mass. Here $\hbar$ is Planck's constant divided by $2\pi$, $c$
is the speed of light in vacuum and $G$ is Newton's universal constant of
gravity. The mass of the $n^{\text{th}}$\ excited state of a Holeum is given
by%
\begin{equation}
m_{H}=2m+\frac{E_{n}}{c^{2}} \label{4}%
\end{equation}
The atomic transitions of a Holeum will give rise to gravitational radiation,
whose frequencies have been predicted in \cite{Chavda.Chavda}. Since the
Holeum emits only gravitational radiation and no electromagnetic radiation, it
is a dark matter candidate. The condition for the stability of a Holeum is%
\begin{equation}
m<m_{c} \label{5}%
\end{equation}
where $m$ is the mass of the two identical primordial black holes that
constitute the bound state and%
\begin{equation}
m_{c}=0.8862m_{P}=1.0821\times10^{19}\text{ GeV} \label{6}%
\end{equation}
The radius of the of the $n^{\text{th}}$\ excited state of a Holeum is given
by%
\begin{equation}
r_{n}=\left(  \frac{n^{2}R}{\alpha_{g}^{2}}\right)  \left(  \frac{\pi^{2}}%
{8}\right)  \label{7}%
\end{equation}
where $R=\left(  \frac{2mG}{c^{2}}\right)  $ is the Schwarzschild radius of
the two identical primordial black holes that constitute the Holeum. We
present the \textbf{ground state} values of these parameters of Holeums in
Table 1.

\textbf{Table 1. }The ground state values of some parameters of Holeums.%

\begin{tabular}
[t]{|c|c|c|c|}\hline
$m$\textbf{\ (GeV)} & $\left\vert E_{1}\right\vert $\textbf{\ (GeV)} & $r_{1}%
$\textbf{\ (m)} & $m_{H}$\textbf{\ (GeV)}\\\hline
\multicolumn{1}{|l|}{$1.0\times10^{3}$} & \multicolumn{1}{|l|}{$1.124437\times
10^{-62}$} & \multicolumn{1}{|l|}{$7.233994\times10^{13}$} &
\multicolumn{1}{|l|}{$2.00\times10^{03}$}\\\hline
\multicolumn{1}{|l|}{$1.0\times10^{7}$} & \multicolumn{1}{|l|}{$1.124437\times
10^{-42}$} & \multicolumn{1}{|l|}{$7.233994\times10^{01}$} &
\multicolumn{1}{|l|}{$2.00\times10^{07}$}\\\hline
\multicolumn{1}{|l|}{$1.0\times10^{11}$} &
\multicolumn{1}{|l|}{$1.124437\times10^{-22}$} &
\multicolumn{1}{|l|}{$7.233994\times10^{-11}$} &
\multicolumn{1}{|l|}{$2.00\times10^{11}$}\\\hline
\multicolumn{1}{|l|}{$1.0\times10^{12}$} &
\multicolumn{1}{|l|}{$1.124437\times10^{-17}$} &
\multicolumn{1}{|l|}{$7.233994\times10^{-14}$} &
\multicolumn{1}{|l|}{$2.00\times10^{12}$}\\\hline
\multicolumn{1}{|l|}{$2.5\times10^{12}$} &
\multicolumn{1}{|l|}{$1.098083\times10^{-15}$} &
\multicolumn{1}{|l|}{$4.629756\times10^{-15}$} &
\multicolumn{1}{|l|}{$5.00\times10^{12}$}\\\hline
\multicolumn{1}{|l|}{$5.0\times10^{12}$} &
\multicolumn{1}{|l|}{$3.513865\times10^{-14}$} &
\multicolumn{1}{|l|}{$5.787195\times10^{-16}$} &
\multicolumn{1}{|l|}{$1.00\times10^{13}$}\\\hline
\multicolumn{1}{|l|}{$7.5\times10^{12}$} &
\multicolumn{1}{|l|}{$2.668341\times10^{-13}$} &
\multicolumn{1}{|l|}{$1.714725\times10^{-16}$} &
\multicolumn{1}{|l|}{$1.50\times10^{13}$}\\\hline
\multicolumn{1}{|l|}{$1.0\times10^{13}$} &
\multicolumn{1}{|l|}{$1.124437\times10^{-12}$} &
\multicolumn{1}{|l|}{$7.233994\times10^{-17}$} &
\multicolumn{1}{|l|}{$2.00\times10^{13}$}\\\hline
\multicolumn{1}{|l|}{$2.5\times10^{13}$} &
\multicolumn{1}{|l|}{$1.098083\times10^{-10}$} &
\multicolumn{1}{|l|}{$4.629756\times10^{-18}$} &
\multicolumn{1}{|l|}{$5.00\times10^{13}$}\\\hline
\multicolumn{1}{|l|}{$5.0\times10^{13}$} &
\multicolumn{1}{|l|}{$3.513865\times10^{-09}$} &
\multicolumn{1}{|l|}{$5.787195\times10^{-19}$} &
\multicolumn{1}{|l|}{$1.00\times10^{14}$}\\\hline
\multicolumn{1}{|l|}{$7.5\times10^{13}$} &
\multicolumn{1}{|l|}{$2.668341\times10^{-08}$} &
\multicolumn{1}{|l|}{$1.714725\times10^{-19}$} &
\multicolumn{1}{|l|}{$1.50\times10^{14}$}\\\hline
\multicolumn{1}{|l|}{$1.0\times10^{14}$} &
\multicolumn{1}{|l|}{$1.124437\times10^{-07}$} &
\multicolumn{1}{|l|}{$7.233994\times10^{-20}$} &
\multicolumn{1}{|l|}{$2.00\times10^{14}$}\\\hline
\multicolumn{1}{|l|}{$2.5\times10^{14}$} &
\multicolumn{1}{|l|}{$1.098083\times10^{-05}$} &
\multicolumn{1}{|l|}{$4.629756\times10^{-21}$} &
\multicolumn{1}{|l|}{$5.00\times10^{14}$}\\\hline
\multicolumn{1}{|l|}{$5.0\times10^{14}$} &
\multicolumn{1}{|l|}{$3.513865\times10^{-04}$} &
\multicolumn{1}{|l|}{$5.787195\times10^{-22}$} &
\multicolumn{1}{|l|}{$1.00\times10^{15}$}\\\hline
\multicolumn{1}{|l|}{$7.5\times10^{14}$} &
\multicolumn{1}{|l|}{$2.668341\times10^{-03}$} &
\multicolumn{1}{|l|}{$1.714725\times10^{-22}$} &
\multicolumn{1}{|l|}{$1.50\times10^{15}$}\\\hline
\multicolumn{1}{|l|}{$1.0\times10^{15}$} &
\multicolumn{1}{|l|}{$1.124437\times10^{-02}$} &
\multicolumn{1}{|l|}{$7.233994\times10^{-23}$} &
\multicolumn{1}{|l|}{$2.00\times10^{15}$}\\\hline
\multicolumn{1}{|l|}{$1.0\times10^{17}$} &
\multicolumn{1}{|l|}{$1.124437\times10^{08}$} &
\multicolumn{1}{|l|}{$7.233994\times10^{-29}$} &
\multicolumn{1}{|l|}{$2.00\times10^{17}$}\\\hline
\multicolumn{1}{|l|}{$1.0\times10^{18}$} &
\multicolumn{1}{|l|}{$1.124437\times10^{13}$} &
\multicolumn{1}{|l|}{$7.233994\times10^{-32}$} &
\multicolumn{1}{|l|}{$2.000011\times10^{18}$}\\\hline
\multicolumn{1}{|l|}{$1.0\times10^{19}$} &
\multicolumn{1}{|l|}{$1.124437\times10^{18}$} &
\multicolumn{1}{|l|}{$7.233994\times10^{-35}$} &
\multicolumn{1}{|l|}{$2.112444\times10^{19}$}\\\hline
\end{tabular}

It can be readily seen that the physical properties of Holeums of masses
ranging between $10^{13}$ GeV\ and $10^{14}$ GeV make them promising
candidates for a super heavy dark matter (SHDM) X-particle. A Holeum is the
gravitational analogue of a hydrogen atom. It is as stable as the latter. It
is well-known that hydrogen undergoes pressure ionization in concentrations of
the gas such as stars, galaxies, nebulae, etc. Holeums too can undergo
pressure ionization via a similar process. Large concentrations of Holeums in
the galactic halos may eventually lead to structure formation such as Holeum
stars and their clusters. The existance of a dark matter galaxy in the Virgo
cluster has been reported \cite{R.F.Minchin.et.al}. The temperatures that
exist in the interiors of stars of ordinary matter may also be available in
such large concentrations of Holeums referred to above.

From Table 1 we can see that the ionization energy of a Holeum having
constituent masses $10^{15}$ GeV or greater is $11.24$ MeV or greater which
corresponds to a temperature greater than $10^{11}$ K that is far higher than
that available in stars and galaxies. The ionization energy of a Holeum
consisting of two identical PBHs of mass $10^{14}$ GeV is $112.4$ eV. This
corresponds to a temperature of $1.305\times10^{6}$ K which readily obtains in
stars. It is clear that the pressure ionization of Holeums having constituent
masses around $10^{14}$ GeV is possible in stars. On the other hand, a Holeum
of mass of the order of $10^{13}$ GeV has an ionization energy of the order of
$10^{-3}$ eV, which corresponds to a temperature of the order of $10$ K -
which is almost the same as the present day CMBR temperature. Such Holeums
would ionize easily and would not be expected to survive in the universe
today. We can therefore expect Holeums of masses $\sim$ $10^{13}$\ $-$
$10^{14}$ GeV\ and above to be stable enough to survive in the present-day universe.

Because of the possible copious production of PBH in the early universe
\cite{Ya.B.
Zeldovich,S.W.Hawking4,M.Yu.Khlopov.et.al,A.G.Polnarev.et.al,S.W.Hawking5,J.Garriga.et.al,R.Caldwell.et.al,J.H.MacGibbon.et.al,M.Crawford.et.al,S.W.Hawking6,H.Kodama.et.al,D.La.et.al,I.G.Moss.et.al,H.Kodama.et.al.2,K.Maeda.et.al,R.V.Konoplich.et.al,M.Yu.Khlopov.et.al.2,V.I.Dokuchaev.et.al,S.G.Rubin.et.al,S.G.Rubin.et.al.2,M.Yu.Khlopov.et.al.3}%
, Holeums could be an important component of dark matter in the universe
today. Holeums interact only gravitationally. By the time the particles of
ordinary matter such as atoms and molecules were formed in the early universe,
the gravitational interaction had become the weakest. In addition to the
gravitational interaction, the particles of ordinary matter have three other,
stronger, interactions. This would readily result in a segregation of Holeums
from the particles of ordinary matter. But due to gravity the former would
still cling to the latter, very weakly. This therefore would have resulted in
the accumulation of Holeums in the halos of galaxies. Another reason for the
accumulation of Holeums in the halos of galaxies is the centrifugal force of
the rotating galaxies. Holeums are generally much more massive than the
particles of ordinary matter. Therefore the centrifugal force on them is
greater, and this would naturally pull Holeums to the fringes and the halos of
galaxies. Since Holeums are invisible, so are the galactic halos. Holeums of
masses $\sim$ $10^{13}$ $-$ $10^{14}$ GeV and above would therefore be an
important constituent of the SHDM present in the galactic halos. Holeums of
masses between $10^{13}$ $-$ $10^{14}$ GeV would occasionally undergo pressure
ionization due to local conditions in galactic halos. Holeums of masses $\sim$
$10^{13}$ $-$ $10^{14}$ GeV and above can therefore be considered promising
candidates for the SHDM X-particle.

\section{Holeum origin of UHECR}

The ionization of a Holeum will destroy the bound state and expose the two
individual black holes that constitute the Holeum. These will undergo Hawking
evaporation and explode instantly into two cascades of particles, and $\gamma
$-rays. Some of the ejecta from the black hole evaporation will hadronize into
ultra-high energy particles. Some of the particles emitted can carry kinetic
energies of the order of $1\%$ of the total energy emitted \cite{C.Barbot}.
These are the UHECR observed on the earth. Each burst of Hawking radiation
will have a total energy of upto $10^{23}$ eV. Remembering that a part of the
energy liberated will be in the form of the rest energies of the emitted
particles, the figure $10^{20}$ eV for the highest energy of the cosmic rays
observed on earth is readily comprehensible. This mechanism can explain cosmic
rays of all observed energies. Holeums can therefore be identified as the SHDM
X-particles that are thought to be the sources of UHECR
\cite{C.Barbot,C.Barbot.et.al,C.Barbot.et.al.2}. In this theory of the origin
of UHECR, the particles are created with such high energies that the need for
exotic sources of acceleration does not arise.

Holeums of masses higher than $10^{14}$ GeV would be extremely stable and
their decays would consequently be very rare. Such decays, however, would
liberate energies of upto the order of $10^{28}$ eV which correspond to
Holeums having masses of the order of $1\times10^{19}$ GeV. Such Hyper High
Energy Cosmic Rays (HHECR) would be direct evidence of the decay of heavy
Holeums. Thus, Holeums can give rise to cosmic rays in the energy range upto
the order of $10^{28}$ eV of which only those upto $10^{21}$ eV have so far
been observed.

\section{UHECR from the galactic halo and the absence of the GZK cut-off}

A unique property of Holeums is that they tend to accumulate almost
exclusively in the galactic halos. Thus, we make the unique prediction that
\emph{UHECR will originate mainly in the galactic halos}. The non-central
position of the sun in the galactic halo should therefore result in an
anisotropic flux of UHECR observed on earth
\cite{S.L.Dubovsky.et.al,R.Aloisio.et.al}. There is observational support for
this prediction. As far back as 1983, Giler observed: "The observed anisotropy
can be accounted for only if the diffusion in the disk is much smaller than
that in the halo" \cite{M.Giler}. This property of Holeums provides a natural
explanation for the absence of the GZK cut-off. As discussed above, the GZK
cut-off significantly affects those cosmic rays that originate at large
cosmological distances ($\gg$ $100$ Mpc) from the earth, but not those whose
sources are nearby, within the local supercluster. According to the Holeum
origin model, cosmic rays arise due to the pressure-ionization of Holeums in
the halos of galaxies. It is obvious that the number of the cosmic rays
emitted by the Holeums accumulated in the local galactic halo will be
overwhelmingly large compared to that of those originating at cosmological
distances. Any depletion caused by the GZK cut-off affects only the latter
class which is insignificant in quantity. Hence there may not be any
observable depletion in the overall quantity received on the earth. This
provides a plausible explanation for the absence of the GZK cut-off.

\section{Observational Evidence from Cosmic Ray Antimatter}

Further support for the Holeum origin of cosmic rays arises from the
observation of antiparticles, especially antiprotons, in cosmic rays.
Antiprotons are one of the most precisely measured particles among the various
antimatter species in cosmic rays
\cite{G.Badhwar.et.al,C.Shen.et.al,K.Igi.et.al,K.Igi.et.al.2,A.Lionetto.et.al,T.Sanuki.et.al}%
. The Hawking radiation emitted by a black hole contains both particles and
antiparticles but does not conserve the baryon number. Now we can readily see
that a macroscopic stellar-mass black hole does not emit any significant
amount of Hawking radiation. For this we note that the decay rate of a black
hole is given by%
\begin{equation}
\frac{dm}{dt}=-\frac{\kappa}{m^{2}}\label{8}%
\end{equation}
where%
\begin{equation}
\kappa=\frac{g_{\ast}\hbar c^{4}}{7680\pi G^{2}}=9.84197\times10^{73}\text{
GeV}^{4}\label{9}%
\end{equation}
Here $m$ is the mass of the black hole in GeV. The Hawking temperature of a
black hole is given by%
\begin{equation}
T=\frac{hc}{8\pi^{2}k_{B}R}=\frac{hc^{3}}{16\pi^{2}k_{B}Gm}=\frac
{9.45\times10^{35}}{m}\text{ GeV}\label{10}%
\end{equation}
Now from equation (\ref{8}) we can show that the rate of loss of mass by a
solar mass ($1.98892\times10^{30}$ kg. $\equiv$ $1.12\times10^{57}$ GeV) black
hole is equal to $-7.89\times10^{-32}$ eV/s which is negligible because the
black hole will take $10^{32}$ seconds to emit $7.89$ eV of Hawking radiation.
The temperature of such a black hole is of the order of $10^{-9}\,$K; which is
hardly conducive for emission of any type at all. It is therefore evident that
macroscopic black holes cannot produce the antimatter detected in cosmic rays.
It has been theorized that primordial black holes are the source of antimatter
detected on earth \cite{A.Barrau3,A.Barrau4,A.Barrau5}. The theory of Holeum
is in agreement with this theory, and provides a natural way of preserving the
PBHs intact to the present day, until they evaporate due to the ionization of
the Holeum they are part of. The ionization of Holeums, leading to the Hawking
evaporation of its constituent PBHs, therefore provides a natural explanation
for the presence of antimatter in cosmic rays.

\section{Conclusions}

The following are the main features of the theory of the Holeum origin of
cosmic rays:

\begin{enumerate}
\item Holeums of masses $\sim$ $10^{13}$ $-$ $10^{14}$ GeV and above can be
identified as promising candidates for the SHDM X-particle.

\item Pressure ionization of Holeums accumulated in the galactic halos can
give rise to cosmic rays of all observed energies including UHECRs and upto
the order of $10^{28}$ eV.

\item Cosmic rays are a manifestation of the end-stage Hawking radiation burst
of the primordial black holes (PBH) liberated in ionizing collisions among the
stable Holeums in galactic halos.

\item Antimatter detected in cosmic rays could be a signature of their Holeum origin.

\item The galactic halo origin of the UHECR explains the absence of the GZK cut-off.
\end{enumerate}

The Holeum has been conjectured to be a dark matter particle that may be a
component of galactic halos \cite{Chavda.Chavda}. We have shown in this paper
that it provides a natural explanation for a number of cosmological phenomena
related to cosmic rays. It has been theorized that Holeums could give rise to
a class of short lived gamma ray bursts \cite{Al.Dallal}. The theory of
Holeum, though promising, is in the nascent stage and much future theoretical
and observational work needs to be done to prove its adequacy and role in cosmology.

\end{document}